\documentclass[aps,amsmath,amssymb,preprintnumbers,groupedaddress,preprint]{revtex4}
\usepackage{graphicx}
\newcommand{\nc}{\newcommand}
\nc{\beq}{\begin{equation}} \nc{\eeq}{\end{equation}} \nc{\bea}{\begin{eqnarray}}
\nc{\eea}{\end{eqnarray}}

\def\gsim{\mathrel{\rlap{\lower4pt\hbox{\hskip1pt$\sim$}}
    \raise1pt\hbox{$>$}}}       %greater than or approx. symbol

\def\K3{{\bf K3}}

\def\n2d{\cN_{V^*}^{\otimes 2}}

\def\cN{{\mathcal N}}

\def\to{\rightarrow}

\begin{document}

\preprint{UPR-1191-T}
\title{Gravity Trapping on a Finite Thickness  Domain Wall:\\ An Analytic Study
\vspace{0.5in}}

\author{Mirjam Cveti{\v c}$^{1\,, 2}$} \email{cvetic@cvetic.hep.upenn.edu}
\author{Marko Robnik$^{2}$ \vspace{0.3in}}\email{robnik@uni-mb.si}

\vspace{0.5in}

\affiliation{$^{1}$ Department of Physics and Astronomy,
University of Pennsylvania, Philadelphia, USA\\
$^{2}$ CAMTP - Center for Applied Mathematics and Theoretical Physics, University of Maribor, Krekova 2, SI-2000 Maribor, Slovenia
\vspace{1.0in}}

\begin{abstract}

\vspace{0.5in}

\noindent We construct an explicit model of  gravity trapping domain wall potential where for the
first time we can study  explicitly the graviton  wave function fluctuations  for any thickness of
the domain wall. A concrete form of the potential depends on one parameter $0\le x\le
\textstyle{\frac{\pi}{2}}$, which effectively parameterizes the thickness of the domain wall with
specific limits  $x\to 0$ and $x\to \textstyle{\frac{\pi}{2}}$ corresponding  to the thin  and the
thick wall, respectively.  The analysis of the continuum Kaluza Klein fluctuations  yields  explicit
expressions for  both the  small  and large Kaluza Klein energy. We also derive  specific explicit
conditions in the regime $x>1$, for which the fluctuation modes exhibit a resonance behaviour,   and
which could sizably affect the modifications of the four-dimensional Newton's law at distances, which
typically  are by four  orders of magnitude larger than those relevant for  Newton's law
modifications of thin walls.

\end{abstract}

%\pacs{}

\maketitle

\date{today}

\bigskip

\section{Introduction}

The Bogomol'nyi-Prasad-Sommerfied  (BPS) saturated  supergravity  domain walls
\cite{CveticGriffiesRey} which interpolate between two  (BPS)  anti-deSitter vacua possess  the
feature that they can trap gravity \cite{RandallSundrum}. The original infinitely thin, $Z_2$
symmetric domain  wall solution \footnote{For a detailed discussion of
supergravity domain walls in
four-dimensions, including the thin wall limit, see \cite{CveticSoleng,CveticGriffiesSoleng}. Related
thin wall analysis in five- and D-dimensions was provided in dimensions \cite{Kaloper,Kraus} and
\cite{CveticWang}, respectively.} in five-dimensions can be treated explicitly to obtain the
qualitative form of corrections to four-dimensional Newton's law \cite{RandallSundrum}. Subsequent
studies extended some further analysis to features
of the finite thickness  walls
(see \cite{DeWolfeetal,Csakietal} and references therein)  as well as
numerous other subsequent studies of special models,
primarily employing numerical analyses of fixed thickness solutions (for recent studies see e.g.,
\cite{Bazeiaetal} and references therein) \footnote{The study of possible string theory origin,
within the gauged supergravity context, of such walls was initiated in
\cite{BehrndtCvetic,KalloshLinde}.}.

The purpose of this paper is to advance the analysis of the gravity trapping for domain walls in an
explicit analytic model, where the graviton wave function fluctuations can be studied explicitly for
any thickness of the domain wall. The Schr\"odinger potential  depends on one parameter $0\le x\le
\textstyle{\frac{\pi}{2}}$, which effectively parameterizes the thickness
of the domain wall.  The two limits $x\to 0$
and $x\to \textstyle{\frac{\pi}{2}}$ reproduce the infinitely thin and
thick wall limits,respectively. This is the first
analytic explicit example where the graviton wave function can be obtained explicitly for any value
of allowed $x$, both for small and large values of  Kaluza Klein masses of fluctuations.
Intriguingly, we also find the resonance behaviour in the regime $x>1$ and provide an  explicit
analytic study of resonances which in turn can sizably modify the four-dimensional Newton's law at
distances that are typically four orders of magnitude larger than those relevant for modifications in
the thin wall set-up.

The paper is organized in the following way. In section \ref{model} we present the concrete form of
the gravity  potential in the background of  the BPS domain with a finite thickness, interpolating
(in $Z_2$ invariant way) between two  anti-deSitter vacua with negative cosmological constant
$\Lambda$. We discuss in detail the matching conditions on the metric and the explicit form of the
potential that depends only on one free parameter $0\le x \le
\textstyle{\frac{\pi}{2}}$, effectively parameterizing the
thickness of the wall.  In section \ref{wavefunction} we study the graviton wave function
fluctuations and in particular focus on the explicit form of the probability density at the center of
the wall. In subsection \ref{Mrange} we obtain the analytic expressions for the probability density
at the center of the wall both for small and large values of Kaluza Klein  masses. In subsection
\ref{resonances} we analyse quantitatively the conditions under which the wave function exhibits the
resonances. In section \ref{newtonlaw} we study the implications for the modification of Newton's
law, and in particular obtain the  analytic form of Newton's law modifications both  in thin ($x<1$)
and thick ($x>1$) regimes. Conclusions and proposal for further studies are presented  in section
\ref{conclusions}.

\section{Finite Thickness Domain-wall Model}
\label{model}

For the sake of simplicity the model that we shall discuss is chosen to arise from a $Z_2$ symmetric
finite thickness domain wall that interpolates  between two BPS vacua with the negative cosmological
constant $\Lambda$ in five-dimensions \footnote{Studies  of  co-dimension one objects  in other
dimensions with an analogous structure of the potential would also be interesting, but are relegated
to further studies.}. Our motivation is to present an explicit, integrable model of the Schr\"odinger
potential for the graviton wave function modes,  which would be explicitly parameterized by  the
thickness of the wall. Such a model would in turn allow for an explicit  analysis of the graviton
fluctuation modes  and the subsequent analysis of a modification of four-dimensional Newton's law,
thus providing  a unifying, analytic approach  which addresses these
effects as  a   function of
the wall thickness.

We choose the following form of the potential:
\begin{eqnarray} V(z)&=&-V_0\, ; \ \ \ \ \ \ \ \ |z|\le \textstyle{\frac{d}{2}}\, ,\nonumber\\ V(z)
&=&\frac{15}{4(z+\beta)^2}\, ; \ |z|\ge \textstyle{\frac{d}{2}}\, . \label{potential}
\end{eqnarray}
This potential enters  Schr\"odinger  equation for the graviton wave function (for the derivation of
the wave equation and other details  see, e.g., \cite{Csakietal}):

\begin{equation} -\frac{d^2\, \psi_m(z)}{d\,z^2}+V(z)\, \psi_m (z) =m^2\, \psi_m (z) \, ,
\label{schrodinger}
\end{equation}
where the graviton wave function ${\bar h}_{\mu\nu}={\eta}_{\mu\nu} \psi(z)\exp (ik_\mu x^\mu)$ and
$k_\mu k^\mu=m^2$  parameterizes the Kaluza Klein energy  associated with the four-dimensional
momentum $k_\mu$.

The potential $V(z)$ is related to the conformal factor $A(z)$ of the domain-wall metric:
\begin{equation}
d\,s^2\,=\, \exp[-A(z)]\, \left(-d\,t^2\, +\, d\, z^2\, +\, \sum_{i=1}^3\, d\, x_i
^2\right)\label{metric}
\end{equation} through the following relation (see, e.g., \cite{Csakietal}):
\begin{equation}
V(z)\equiv \frac{9}{16}\left(\frac{d\, A(z)}{d\, z}\right)^2-\frac{3}{4}\frac{d^2\,A(z)}{d\,z^2}\, .
\label{VA}
\end{equation}

Along with the boundary conditions  $A(0)=0$, $A(0)'=0$ and $A(z)\to 2\log (k\, |z|)$ as $|z|\to
\infty$ the relation (\ref{VA})  determines for the potential  (\ref{potential}) the following form
of the metric: \begin{eqnarray} A(z)&=&-\frac{4}{3}\log (\cos \sqrt{V_0}|z|)\, ;
\ |z|\le \textstyle{\frac{d}{ 2}}\, , \nonumber\\
A(z) &=&\log [k^2((|z|+\beta)^2]\,  ; \  \ \ \ |z|\ge \textstyle{\frac{d}{2}}\, , \label{metricp}
\end{eqnarray}where $k=\sqrt{-\textstyle{\frac{\Lambda}{6}}}$  (see, e.g.,
\cite{BehrndtCvetic})
 and the five-dimensional Planck constant $M_{Planck_5}$
was set to 1.

The continuity of the metric and its derivative at the junction $|z|=\frac{d}{2}$ imposes the
following two conditions among  parameters $V_0$, $d$ and $\beta$:
\begin{eqnarray} k(\frac{d}{2} +\beta)&=&\cos (\textstyle{\frac{\sqrt{V_0}
d}{2}})^{-{2}/{3}}\, , \nonumber\\
 (\frac{d}{2}+\beta)^{-1} &=&\frac{2}{3}\sqrt{V_0}\tan
(\textstyle{\frac{\sqrt{V_0}d}{2}})\, .
  \label{matching}
\end{eqnarray}
These two conditions  can be viewed as fixing $\textstyle{\frac{d}{2}}+\beta$ and $V_0$ in terms of
{\it one free parameter}: \begin{equation} x\equiv \frac{\sqrt{V_0}{d}}{2}\, .\end{equation} The
parameter $x$ has a range $\{0,\ \textstyle{\frac{\pi} {2}}\}$. The infinitely thin wall corresponds
to the following limit of parameters: \begin{equation}
   x\to 0\, ; \ \ \ d\to 0\, , \ \ \  V_0\to \infty\, , \ \ \ V_0\, d=\frac{3}{\beta}=3k\,
   ,\end{equation}
while the infinitely thick wall corresponds to: \begin{equation} x\to \textstyle{\frac{\pi}{2} }\, ;\
\ \ d\to \infty\, , \ \ \  V_0\to 0\, , \ \ \  \sqrt{V_0}\beta \to - x\, , \ \ \ (\sqrt{V_0}\beta +
x)^5 (\frac{k}{\sqrt{V_0}})^3=\frac{9}{4}\, .
\end{equation}

\section{Wave Function}
\label{wavefunction}

The solution of the Schr\" odinger equation  (\ref{schrodinger}) has eigenvalues  $m^2\ge 0$, with
$m=0$ corresponding to the graviton bound state (see e.g., \cite{Csakietal}):
 \begin{equation} \psi_0(z)=N_0\; e^{ (-\textstyle{\frac{3}{4}}A(z))}\,
,\label{gwf} \end{equation}
 where the normalisation constant $N_0$ is fixed by the following
relationship:
\begin{equation}\frac{k}{N_0^2}=\textstyle{\frac{2}{3}}\sin(x)\cos(x)^{-5/3}\,
 (x+\sin(x)\cos(x))+\cos(x)^{4/3}\, .\label{gwfn}
 \end{equation}
In the thin wall limit  ($x\to 0$)  and thick wall limit ($x\to
\textstyle{\frac{\pi}{2}}$) the normalisation coefficient $N_0$ takes the
following respective forms:
\begin{equation}
\frac{k}{N_0^2} \to 1\, , \ \ {\rm as} \ \ x\to 0\, ; \ \ \frac{k}{N_0^2}\to
\frac{\pi}{3\cos(x)^{5/3}}\, , \ \ {\rm as} \ \ x\to\frac{\pi}{2}\, .
\end{equation}
  The solution
$\psi_m(z)$ of (\ref{schrodinger}) in  the continuum, $m^2>0$, has
the following form:
\begin{eqnarray} \psi_m (z)&=&A_m \cos (K|z|) \, ;\ \ \ \ \ \ \ \ \ \ \ \ \ \ \ \
 \ \   \ |z|\le \textstyle{\frac{d}{2}}\nonumber\\
\psi_m(z) &=&N_m\sqrt{|z|+\beta}\,  [a_m Y_2(m|z|+\beta)+ b_m J_2 (m|z|+\beta)\, ]\, ;
 \ \ \ |z|\ge \textstyle{\frac{d}{2}}\, . \label{wf}
\end{eqnarray}
Here  $K\equiv \sqrt{m^2+V_0}$ and the coefficients $a_m$ and $b_m$ satisfy $a_m^2+b_m^2=1$. The
normalisation constant $N_m$ is determined by employing a regulator $|z_c|\to \infty$ along with the
asymptotic form of $\psi_m(z)$ as $|z|\to\infty$:
\begin{equation}
\psi_m(z)= N_m\sqrt{\textstyle{\frac{2}{\pi m}}}[a_m\, \sin
(m(|z|+\beta)-\textstyle{\frac{5\pi}{4}})+b_m\, \cos (m(|z|+\beta)-\textstyle{\frac{5\pi}{ 4}})]\, .
\label{awf}\end{equation} In the limit $|z_c|\to \infty$ the   asymptotic wave function (\ref{awf})
ensures the dominant contribution to the wave function probability, and thus
$N_m=\sqrt{\textstyle{\frac{\pi\, m } {2\, z_c}}}$, which in the continuum  limit becomes
$N_m=\sqrt{\textstyle{\frac{m}{2}}}$, i.e. $z_c$  is replaced by $\pi$.

The matching of the wave function  and its first derivative at $|z|=\textstyle{\frac{d}{2}}$
determines the coefficients $a_m, \ b_m$ ($a_m^2+b_m^2=1$) and $A_m$. In particular, the key
expression in determining the deviations of  Newton's law is the probability density of the wave
function at the center of the wall:
\begin{equation}|\psi_m(0)|^2=A_m^2 = \frac{1}{\pi^2}\, \frac{ 1} { ({\tilde J}_2 {\tilde S} -{\tilde J}_2'
 {\tilde C})^2  + ({\tilde Y}_2 {\tilde S}-{\tilde Y}_2'
{\tilde C})^2}\, , \label{Ams} \end{equation}
%{\it which is useful for the analysis of the resonance behaviour where both positive definite terms
%in the denominator achieve minimal values for the same M (which happens for  $1.4<x<1.55$).}
where:
 \begin{equation}{\tilde C} = {\cos ({\tilde y}_0)},\;\;\; {\tilde S} = - K \sin ({\tilde y}_0)\, ,  \end{equation}
and
\begin{equation}{\tilde J}_2 \equiv  \sqrt{\frac{y_0}{2}}\, { J}_2 (y_0), \;\;\; {\tilde J}_2'
 \equiv \frac{d\,
{\tilde J}_2(y_0)}{dy_0},\;\;\; {\tilde Y}_2  \equiv  \sqrt{\frac{y_0}{2}}\, {Y}_2(y_0),\;\;\;
{\tilde Y}_2'\equiv \frac{d {\tilde Y}_2(y_0)}{dy_0}\, \end{equation} 
where $J_2$ and $Y_2$ are Bessel functions of order 2.
The arguments $y_0$ and
${\tilde y}_0$ are defined as:
\begin{equation}y_0 = \frac{M }{\cos(x)^{2/ 3}}  \, , \ \ \  \ {\tilde
y}_0 = x\,\sqrt{1
+\frac {4 M^2\sin(x)^2 }{9 \cos(x)^{{10/ 3}}}}\, ,
\end{equation} and coefficients $M$ and $K$ are defined as:

\begin{equation} M\equiv \frac{m}{k}\, , \ \ \ \ K\equiv
 \sqrt{1+\frac{9 \cos(x)^{{10/ 3}}} {4M^2 \sin(x)^2 }}\, . \end{equation}
Note that the expression for $A_m^2$ (\ref{Ams})
 is a function of $x$ and $M$, only!
  This allows us to fully explore the analytic behaviour of the probability density $A_m^2$ which we
  shall do  in the following subsections.
In particular in the thin wall limit ($x\to 0$) $A_m^2$ takes the form:
\begin{equation}
A_m^2=\frac{2}{\pi^2\, M\, [J_1^2(M)+Y_1^2(M)]}\, .
\end{equation}

  For the sake of completeness we also present the expression for the ratio:
  \begin{equation}
  \frac{a_m}{b_m}=-\frac{{\tilde J}_2 {\tilde S} -{\tilde J}_2' {\tilde
C}}{{\tilde Y}_2 {\tilde S}
  -{\tilde Y}_2'
{\tilde C}}\, ,
\end{equation}
and $b_m=1/\sqrt{1+({a_m}/{b_m})^2}$.

 %Therefore everything needed for the calculation of the corrections
%to  Newton's law is expressed in terms of $m, x$ and $k$. So we shall have to evaluate the
%integral

%\begin{equation}\int_0^{\infty} dm \frac{e^{-mr}}{r} \overline{\psi^2} \end{equation}%
%Now we can plot the diagram $A_m$ as a function of $m$ at fixed various values of $x$. Here we might
%discover important "resonances", i.e. large peaks of $A_m$ at certain values of $m >0$.

Note that for the thick wall $x> 1$ the expression for the probability density of the wave function
varies significantly over the range of the wall. In this case the probability density, averaged over
the domain of the wall thickness, may be more appropriate:

\begin{equation} \overline{|\psi_m|^2} = {A_m^2}F_m \, ; \ \ F_m=
{\frac{1}{2}} (1 + \frac{\sin(2{\tilde y}_0)}{2{\tilde y}_0} )\, . \label{awff}\end{equation} Note
that $F_m$ is  a mild function of $x$, and while $F_m\to 1$ as $x\to 0$,
for  $x\to \textstyle{\frac{\pi}{2}}$ it
approaches the asymptotic value $\textstyle{\frac{1}{2}}$.
%In this case it may be of interest to explore a deviation of  Newton's law away from the center of
%the wall by employing the above ``averaged'' probability density.

\subsection{Small and large Kaluza Klein Energy Expansion}\label{Mrange}
The expression (\ref{Ams}) can be readily expanded in different regimes of values of $x$ and $M$
parameters.

The expression (\ref{Ams}) has a universal behaviour: \begin{equation} A_m^2\to \frac{1}{\pi}\, , \ \
{\rm as} \ \ M\to \infty\, . \label{amslM}\end{equation} On the  other hand one can also explicitly
expand (\ref{Ams}) in the limit of small $M$. We have done so up to the fourth order in $M$ expansion:
\begin{equation} A_m^2={M}\alpha_0\,(\,1+ M^2\, \alpha_1\, )+{\cal O}(M^5)\, , \ \ {\rm
for} \ \ M\ll 1\label{amssM}
\end{equation}
where coefficients $\alpha_0$ and $\alpha_1$ are the following explicit functions of $x$:
\begin{equation}
\alpha_0=\frac{{9}\cos (x)^{10/3}}{2(\cos (x)^3\,+\,2\,x\sin( x)\,+\,2\cos (x))^2}\, , \ \ \
\alpha_1=\frac{p_1}{p_2}\, ,\end{equation} where \begin{eqnarray}
 p_1&=& \cos(x)\left( 8- 34 \cos(x)^2 - \cos(x)^4(1 + 54 \log(\frac{2
\cos(x)^{2/3}}{M\exp(\gamma)}))\right)\nonumber\\ &+2&x\sin(x)(4 -9 \cos(x)^2-4\cos(x)^4)\, ,\\
p_2&=&{18 \cos(x)^{10/3}(\cos(x)^3+2x\sin(x)+2\cos(x))}\, .\label{amssMp}
\end{eqnarray}
The thin wall limit ($x\to 0$) produces the following values of $\alpha_0$
and $\alpha_1$:
\begin{equation}
\alpha_0=\frac{1}{2}\, , \ \
\alpha_1=-\textstyle{\frac{1}{2}}+\log(\textstyle{\frac{M\exp(\gamma)}{2}})\, ,\ \ {\rm as}\ \ x\to
0\, .\end{equation}  Note  (\ref{amssMp}) is a valid expansion for $x<\textstyle{\frac{\pi}{2}}$,
however in the thick wall limit ($x\to\textstyle{\frac{\pi}{2}}$), $A_m$ has further suppressions for
small values of $M$:
\begin{equation}
 \alpha_0\to \textstyle{\frac{9}{2\pi^2}} ({\textstyle{\frac{\pi}{2}}}-x)^{10/3}\, , \
 \alpha_1\to \textstyle{\frac{2}{9}}({\textstyle{\frac{\pi}{2}}}-x)^{-10/3}\,, \ \ {\rm
as} \ \ x\to \frac{\pi}{2} . \end{equation} and thus $A_m^2\to \textstyle{\frac{1}{\pi^2}}M^3$.

\begin{figure}
\begin{center} \scalebox{0.55}{{\includegraphics[angle=270]{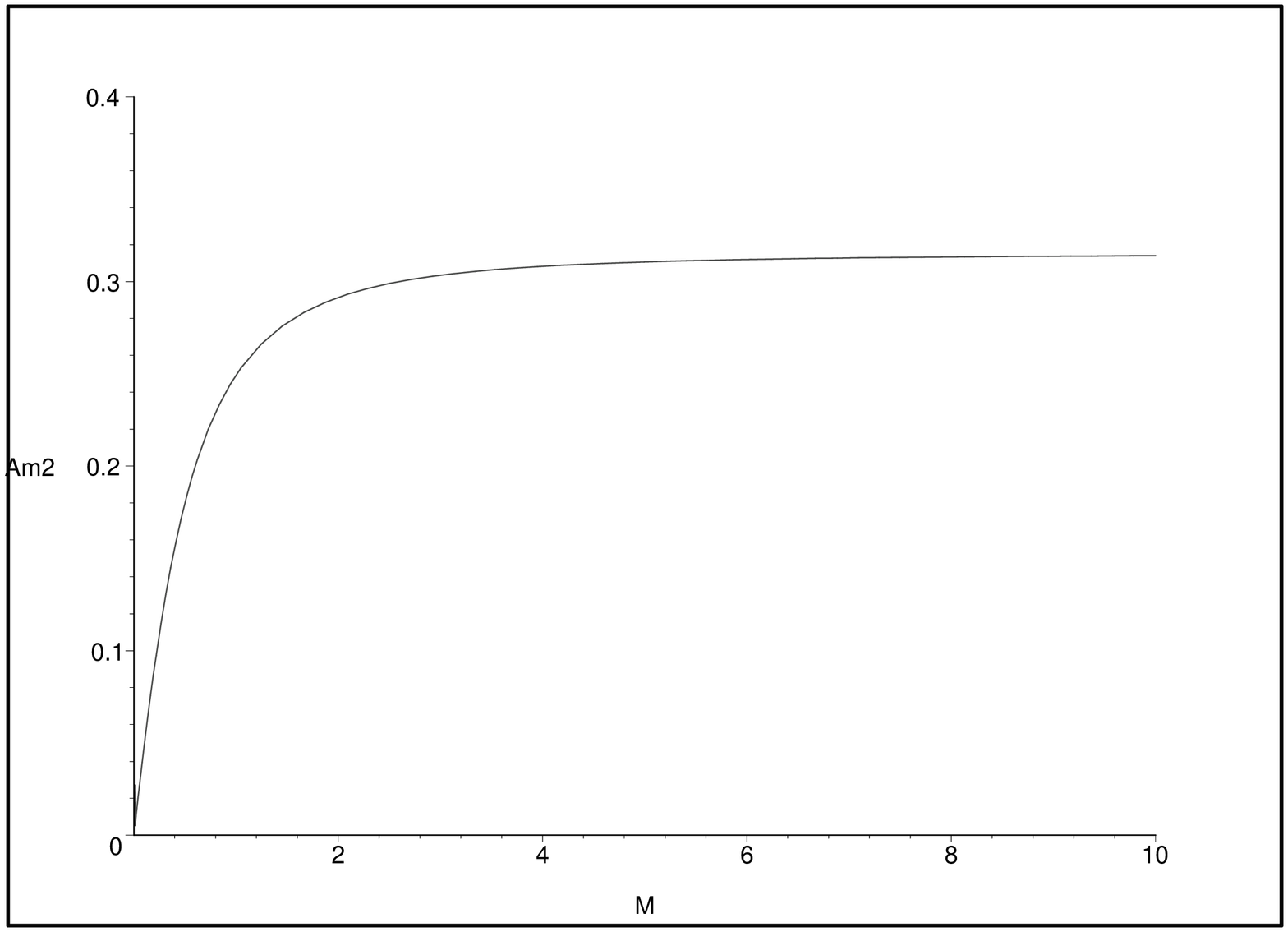}}}
%\centering \epsfysize=10cm \leavevmode \epsfbox{MajorO1.eps}
\end{center}
\caption[]{\small Probability density $A_m^2$  plotted as a function  of M for a small value of
$x=0.1$.
 }
 \label{generic}
\end{figure}

\begin{figure}
\begin{center}
\scalebox{0.55}{{\includegraphics[angle=270]{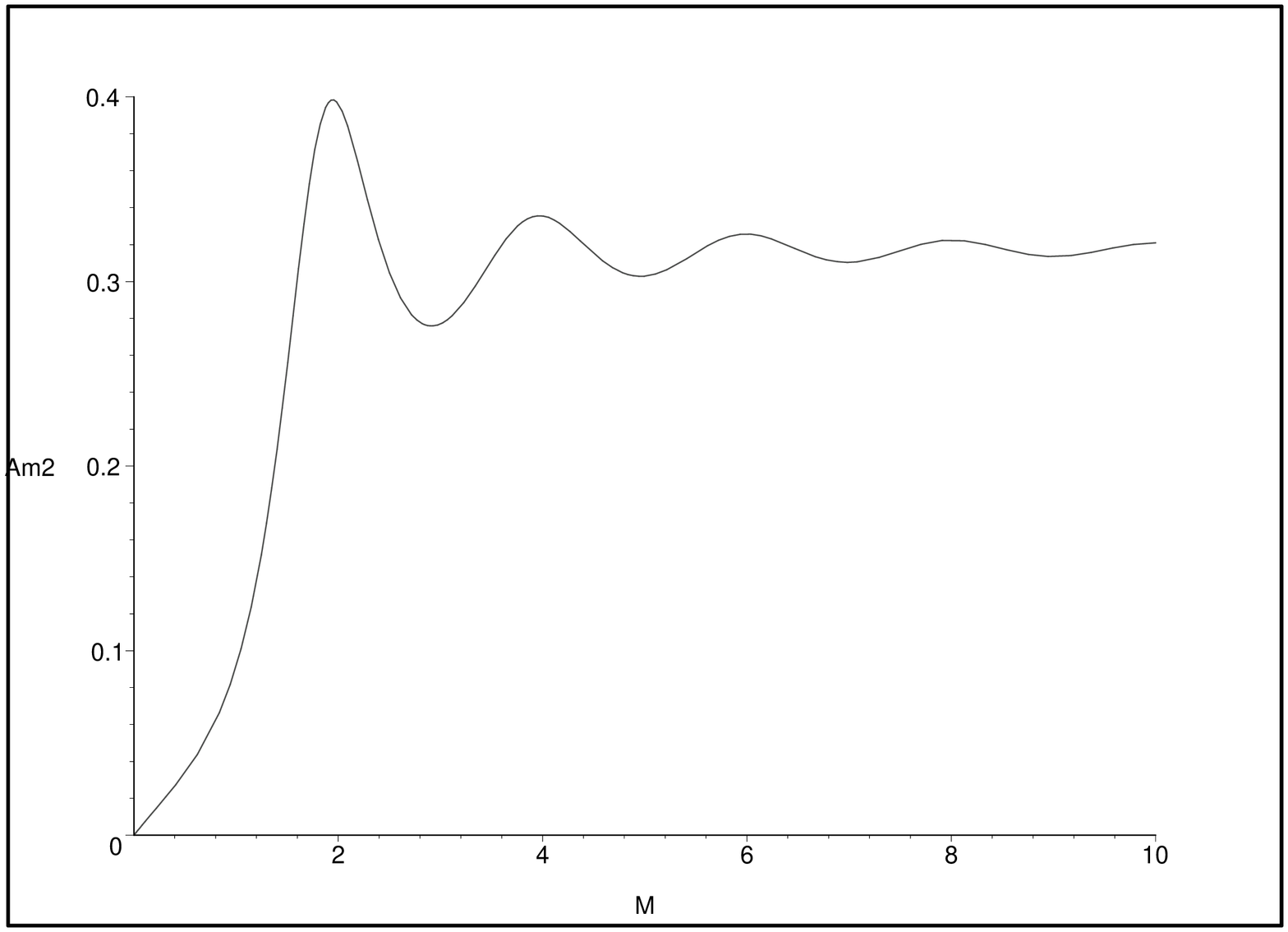}}}
%\centering \epsfysize=10cm \leavevmode \epsfbox{MajorO1.eps}
\end{center}
\caption[]{\small Probability density $A_m^2$  plotted as a function  of M for $x=1$.
 }
 \label{genericp}
\end{figure}

 In figure
\ref{generic} and \ref{genericp} $A_m^2$ is plotted for a small and an intermediate value of $x$,
respectively.

For the sake of completeness we also give the expansion:
\begin{equation} \frac{a_m}{b_m}={M}^2\beta_0\,(\,1+
M^2\, \beta_1\, )+{\cal O}(M^6)\, , \ \ {\rm for} \ \ M\ll 1\, ,
\end{equation}
where:
\begin{equation}
\beta_0=\frac{3\pi \cos(x)^{5/3}}{4(\cos(x)^3+2x\sin(x)+2\cos(x))}\, , \ \ \
\beta_1=\frac{r_1}{r_2}\, , \end{equation} where
\begin{eqnarray}
r_1&=&4x\sin(x)\cos(x)(2-11\cos(x)^2)+\cos(x)^2(28-92\cos(x)^2-17\cos(x)^4) \nonumber\\
&-&4x^2\sin(x)^2(5-8\cos(x)^2)-108\cos(x)^6\log(\frac{2 \cos(x)^{2/3}}{M\exp(\gamma)})\, ,
\\r_2&=&{72\cos(x)^{13/3}(\cos(x)^3+2x\sin(x)+2\cos(x))}\, \end{eqnarray}
whose thin and thick limits have the following form:
\begin{eqnarray}
\beta_0=\textstyle{\frac{\pi}{4}} \, &,& \ \ 
\beta_1=-\textstyle{\frac{3}{8}}+\textstyle{\frac{1}{2}}\log(\frac{M\exp(\gamma)}{2})\, , \  \ {\rm as} \ \  x\to 0;\nonumber\\
\beta_0\to \textstyle{\frac{3}{4}} ({\textstyle{\frac{\pi}{2}}}-x)^{5/3}\, &,& \beta_1\to
-\textstyle{\frac{5\pi}{72}}({\textstyle{\frac{\pi}{2}}}-x)^{-13/3}\,, \ \ \ \ \ \ \ \ \  {\rm as} \
\ x\to \frac{\pi}{2} .
\end{eqnarray} 

\subsection{Resonances}
\label{resonances} 

Another interesting behaviour of $A_m$
takes place in a particular window for a combination of parameter $x>1$ and small values of $M$:
\begin{equation}
y_0 = \frac{M}{\cos(x)^{{2/ 3}}}\ll 1 \, , \ \ \ {\tilde y}_0 = x\,\sqrt{1
+\frac {4M^2 \sin(x)^2 }{9
\cos(x)^{{10/ 3}}}}\gg 1\, , \label{reslim}
\end{equation}
or
\begin{equation}
\frac{M} {\cos(x)^{2/ 3}}\ll 1\, , \ \ \  \frac{M\tan(x)} {\cos(x)^{2/ 3}}\gg 1\,.\label{reslimp}
\end{equation}
The  probability density  (\ref{Ams}) exhibits a resonance  behaviour
\footnote{The appearance of
resonances is analogous to that of a simpler set-up with  a square-well potential, where however both
the depth and the thickness are free parameters. In our case it is  remarkable  that even though we
have only one free parameter $x$, that  specifies the potential, nevertheless the resonances can
appear!}.
 The resonance then appears when ${\tilde Y}_2{\tilde S}-{\tilde Y}_2' {\tilde C}\sim 0$,
  which for  the
regime (\ref{reslimp}) takes the form: \begin{equation} \frac{\tilde S}{\tilde C}\sim
-\frac{3}{2y_0}\, . \label{rescon}
\end{equation}
For  the above expression we have employed the small argument $y_0\ll 1$ expansion of ${\tilde Y}_2=
-\textstyle{\frac{2\sqrt 2}{\pi y_0^{3/2}}} +{\cal O}(y_0^{1/2}) $ and ${\tilde Y}_2'=
\textstyle{\frac{3\sqrt 2}{\pi y_0^{5/2}}}+{\cal O}(y_0^{-1/2}) $.  Eq. (\ref{rescon}) is satisfied
approximately when $\tilde y_0\sim \textstyle{\frac{\pi}{2}}(2n+1)$ ($n=1,2,\cdots$) or:
\begin{equation} M_n\sim  3\cos(x)^{5/3}\, \sqrt{n(n+1)}\, , \ \ \ n=1,\ 2,\ \cdots \, .\label{Md}\end{equation}
Note that the above relation for the position of resonances becomes less valid  as $n$ increases.
%Note that for  condition ${2M\tan(x)}/ ({3\cos(x)^{2/ 3}})\sim \ 1$,   ${\tilde y_0}\sim
%\pi/\sqrt{2}$ with $\frac{\tilde S}{\tilde C}>0$ and thus it is not compatible with the resonance
%condition (\ref{rescon}).

 When  the condition (\ref{rescon}) is satisfied, the probability density (\ref{Ams}) is
highly peaked at
\begin{equation} A_m^2\sim  \frac{18}{\pi^2\, y_0^5}
= \frac{18\cos(x)^{10/3}}{\pi^2\,M_n^5}\gg 1\, .\label{resamp}\end{equation}  To obtain the above
expression,  we set in (\ref{Ams}):  ${\tilde Y}_2{\tilde S}-{\tilde Y}_2' {\tilde C}\sim  0$, and
employed the small $y_0$ argument expansion of
 of ${\tilde J}_2=\textstyle{\frac{
(\sqrt 2\, y_0^{5/2})}{16}} +{\cal O}(y_0^{7/2}) $ and ${\tilde J}_2'= \textstyle{\frac{5\sqrt 2\,
y_0^{3/2}}{32}}+{\cal O}(y_0^{5/2}) $  which yields (along with (\ref{rescon})): ${\tilde J}_2 {\tilde
S} -{\tilde J}_2'
 {\tilde C} \sim  -\textstyle{\frac{y_0^{5/2}}{3\sqrt{2}}}$. Note the sharp fall-off of the
density amplitude with the increase of the discretely valued $M_n$ (\ref{Md}).

 One can also estimate the
approximate width $\Delta M_n$ of the resonances,
 by determining the change in the parameter $M$ (by
$\textstyle{\frac{1}{2}}\Delta M_n$) when ${\tilde Y}_2{\tilde S}-{\tilde
Y}_2' {\tilde C}\sim  {\tilde J}_2 {\tilde S}
-{\tilde J}_2'
 {\tilde C}\sim  -\textstyle{\frac{y_0^{5/2}}{3\sqrt{2}}}$,
using locally linear approximation. 
This condition yields:
\begin{equation} \frac{\Delta M_n}{M_n}\sim  \frac{\pi y_0^5}{9({\tilde
y}_0-\frac{\pi^2}{4{\tilde y_0}})}=
\frac{M_n^4}{3\cos(x)^{5/3}}\sqrt{1+\textstyle{\frac{1}{4n(n+1)}}}\, . \label{width}
\end{equation}
Note that in this case the estimate for $A_m^2\, \Delta M_n$, which
effectively parameterizes the
integral of the  sharply peaked density probability over its width, takes the form:
\begin{equation}
A_m^2\, \Delta M_n\sim  \frac{2 M_n}{\pi ({\tilde y}_0-\frac{\pi^2}{4{\tilde y_0}})}=
\frac{6\cos(x)^{5/3}}{\pi^2}\sqrt{1+\textstyle{\frac{1}{4n(n+1)}}}\, , \label{resint}
\end{equation}
and is thus close to being  $M_n$ independent! Of course, this is 
a valid approximation only for the
first few resonances whose width is still small, as we used only locally linear
approximation for the quantity $\tilde{Y_2}\tilde{S} - 
\tilde{Y_2'}\tilde{C}$ near $M_n$.

 \begin{figure}
\begin{center}
\scalebox{0.55}{{\includegraphics[angle=270]{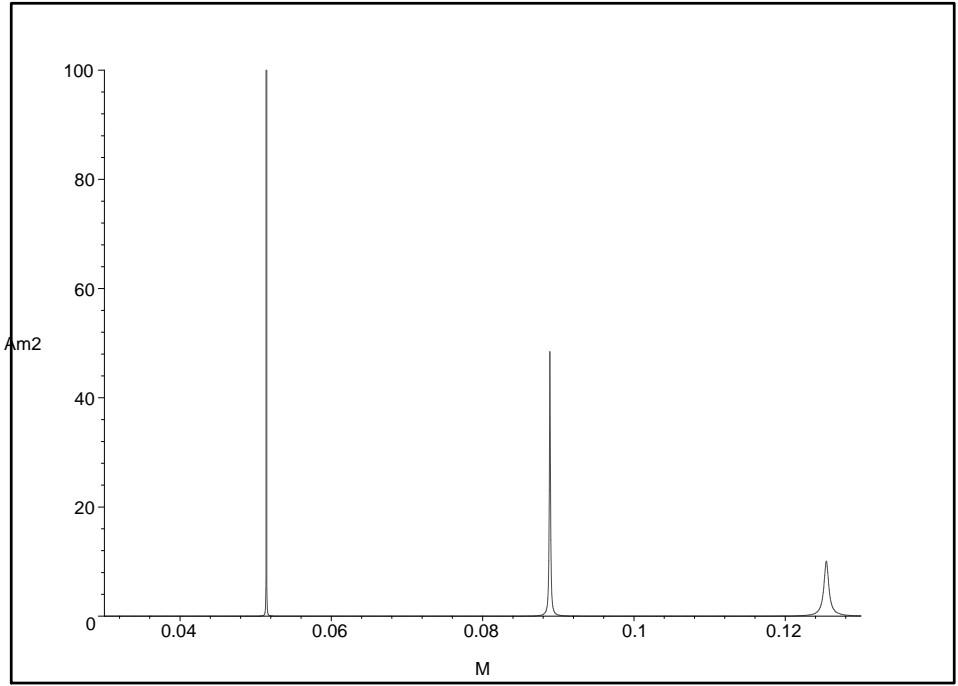}}}
%\centering \epsfysize=10cm \leavevmode \epsfbox{MajorO1.eps}
\end{center}
\caption[]{\small Probability density $A_m^2$  plotted as a function  of M for the  value of $x=1.5$.
 It exhibits distinct resonances.}
%
%The location of O6-orientifold fixed planes
%and their three  $\IZ_2\times \IZ_2$  images  (denoted by  bold solid lines)
%for the case of a six-torus factorized on  three rectangular
%two-tori.  }
\label{reson}
\end{figure}
 The most distinct and pronounced resonances  appear in the range  $x\sim\{1.35,1.55\}$.
 In figure
\ref{reson} the density probability as a function of M is plotted for the regimes of $x$
 where the resonances are pronounced. We have checked explicitly
  that our analytic
 results for the  location, the amplitude and the width of resonances are in good
  agreement with numeric ones, as it is evident from table \ref{annum}.
  \renewcommand{\arraystretch}{1.4}
\begin{table}[t]

\begin{center}
\begin{tabular}{||c||c||c|c|c||}
\hline\hline$\ $ &
 n &1&2&3\\\hline\hline
 ${\rm numeric}$&$M_n$&0.05141&0.08889& 0.12543  \\ \hline
 ${\rm analytic}$ &$M_n$&0.05133 &0.08891  & 0.12574  \\ \hline\hline
  ${\rm numeric}$&$A_m^2$&680&50.07&10.12  \\ \hline
  ${\rm analytic}$ &$A_m^2$&749 &48.06 & 8.50 \\ \hline\hline
  ${\rm numeric}$&$\Delta M_n$&$1.07\, 10^{-5}$&$1.47\, 10^{-4}$&$7.16\,
10^{-4}$  \\ \hline
  ${\rm analytic}$&$\Delta M_n$&$ 1.04\, 10^{-5}$&$ 1.56\,  10^{-4}$ &
$8.75\, 10^{-4}$ \\ \hline\hline
\end{tabular}
\end{center}
\vspace{0.2cm}\caption{The comparison of the numeric and analytic results for the location $M$, the
amplitude $A_m^2$ and the width  $\Delta M$ of the resonances for $x=1.5$.}
\vspace{0.3cm}\label{annum}
\end{table}
%In particular, for $x=1.5$, the resonances at ($n=1,2,3$) yield the following numeric and analytic
%results:
% \begin{eqnarray}   {\rm numeric}: \ M&=&(0.0514,\, 0.0889,\, 0.1254)\, ; \ A_m^2= (680,\,50,\, 10)\,; \
%\Delta M=(1.210^{-5},\, 1.410^{-4},\, 8 10^{-4}\, ;\nonumber \\
%  {\rm analytic}:\ M&=&(0.054,\, 0.0907,\ ,0.127)\, ; \ A_m^2= (560,\, 43,\, 8)\,; \
%\Delta M=(310^{-5},\, 510^{-4},\, 3 10^{-3})\,.\end{eqnarray}

\section{Modification of Newton's Law}\label{newtonlaw}
The analytic form of the probability density in turn allows us study explicitly deviation of Newton's
law in the presence of the continuum states. The (bound state) graviton wave function (\ref{gwf})
determines the four-dimensional  Newton's law, while continuum Kaluza Klein fluctuation modes
(\ref{wf}) are responsible for corrections to  Newton's law. The four-dimensional gravitational
potential between two point-like particles with $M_1$ and $M_2$, a distance $r$ apart,  can be
written in the form (see e.g., \cite{Csakietal}):
\begin{equation}
V=G_N\frac{M_1M_2}{r}(1+ \delta)\, , \end{equation} where $G_N=M_{Planck_5}^{-3}N_0^2$ and
$\delta$,  the correction to Newton's law is of the form:
\begin{equation}
\delta=\frac{1}{N_0^2}\int_{0^+}^{\infty} e^{-mr}
\overline{|\psi_m|^2}\,dm\, \, .\label{nlc}
\end{equation}
%where $G_N=M_{Planck5}^{-3}N_0^2$.
%\int_{-\infty}^{\infty}\,  e^{-\frac{3 A(z)}{2}}\, dz$ {\bf check}.
Note that since  $N_0^2/k$ (eq. (\ref{gwfn})) is only a function of $x$
and   $\overline{|\psi_m|^2}$  (eq.(\ref{awff}))
is only a function of $x$ and $M\equiv m/k$,
the correction to Newton's law can be rewritten as an integral
over $M$:
\begin{equation}
\delta=\frac{k}{N_0^2}\int_{0^+}^{\infty} e^{(-M\, kr)}
\overline{|\psi_m|^2}\,dM, \, \label{nl}
\end{equation} and it is thus  only a function of $kr$ and $x$!

Note again that the averaged wave function density $\overline{|\psi_m|^2}$ (eq.(\ref{awff}))  is of
the form $A_m^2 F_m$ where $F_m$ is a mild function of $x$ and $M$, with $F_m\to 1$ and $F_m\to 1/2$
as $x\to 0$ and $x\to \textstyle{\frac{\pi}{2}}$, respectively.  In the
following we shall  focus only on $A_m^2$ factor
and its effect on the structure of $\delta$.

The amplitude   $A_m^2$ (eq.(\ref{Ams})) has a small
$M$ expansion of the form (\ref{amssM}) and (\ref{amssMp})
 and it asymptotes to $\pi^{-1}$ as $M\to \infty$ ((eq.\ref{amslM})). For
$x\le 1$ it is a reasonable
 approximation to split the integral (\ref{nl}) into  two intervals
$M=\{0,M_C\}$ and $M=\{M_C,\infty\}$
 where the form of $A_m^2$ the small $M$ and large $M$ expansion is used, respectively. A reasonable
 choice $M_C\sim (\pi \alpha_0)^{-1}$, produces an approximate result which is in good agreement with numerical
 one:
 \begin{eqnarray}
 \delta&=&\frac{k}{N_0^2} \{\frac{\alpha_0}{(kr)^2}[1-e^{(-M_C\, kr)})(M_C\,kr+1)]\nonumber\\
 &+&\frac{\alpha_0\alpha_1}{(kr)^4}[6-e^{(-M_C\, kr)}((M_Ckr)^3+3(M_Ckr)^2+6M_C\,kr+6)]
 +\frac{1}{\pi kr}
  e^{(-M_C\, kr)}\}\, .\label{xl1c}\end{eqnarray}
 Again, note for $r\gg (kM_C)^{-1}$ the leading correction is of the form:
 \begin{equation} \label{deltalarger}
 \delta=\frac{k}{N_0^2} \frac{\alpha_0}{(kr)^2}(1+
 \frac{6\alpha_1}{(kr)^2})
 \, . \end{equation}

 The intriguing possibility takes place for $x> 1$ where the resonance behaviour takes place.
 Note that in this case the resonances are highly peaked, and the integral over $M$ can be approximated
 by a sum of expressions  $kN_0^{-2}\, A_m^2\Delta M_n e^{-M_nkr}$  over $n$.
 Employing the analytic expressions  from subsection \ref{resonances} we
obtain the following
correction to Newton's law:
\begin{equation}
\delta\sim  \frac{2}{\pi}\sum_n \sqrt{1+\textstyle{\frac{1}{4n(n+1)}}}\, e^{-M_n\, kr}\, ,
\end{equation} where $M_n$ are locations of
resonances (\ref{Md})  and the approximately constant prefactor is obtained by using the limit $x\to
\textstyle{\frac{\pi}{2}}$ in (\ref{gwfn}):  $kN_0^{-2}\to \textstyle{\frac{\pi}{3}}\cos(x)^{5/3}$
and (\ref{resint}): $(A_{m}^2\Delta M_n)\sim
\textstyle{\frac{6}{\pi^2}}\cos(x)^{5/3}\sqrt{1+\textstyle{\frac{1}{4n(n+1)}}}$. Note that the
prefactor is approximately constant and of order one.

$M_n$'s (\ref{Md}) are typically in the range of $10^{-2}$ and thus for a choice of also small k,
 and
 the range of distanced $r\sim (M_nk)^{-1}$ the leading resonances can
potentially
 contribute a sizable effect!
In particular when  choosing $k\sim 10^6\, {\rm GeV}$, at distances that
 $r\sim 10^{-4}{\rm GeV}^{-1}$ the thick wall limit with
resonances $M_n\sim 10^{-2}$ would produce an order one modification of  Newton's law, while for
the walls with $x\le 1$  the corrections (\ref{xl1c},\ref{deltalarger}) would be negligible, i.e.  ${\cal O}(10^{-4})$.

 \section{Conclusions}\label{conclusions}
We have presented the first explicit model of a finite thickness domain wall, interpolating between
$Z_2$  symmetric  five-dimensional anti-deSitter vacua, where the graviton wave function fluctuations
can be studied explicitly for any thickness of the wall, parameterized by a parameter
$x=\{0,\textstyle{\frac{\pi}{2}}\}$. This allows us to determine
explicitly the probability density of the Kaluza Klein
fluctuations both for the small and large values of the Kaluza Klein energy. Notably, for $x>1$
resonance behaviour emerges, which is most pronounced in the range
$x=\{1.35,1.55\}$ and can be
analysed explicitly. For a specific range of Kaluza Klein momenta (and anti-deSitter cosmological
constant $\Lambda$) this resonance behaviour can significantly modify four-dimensional Newton's law
by an effect of order  $10^4$ larger than that of thin walls.

While the concrete model is based on a specific, analytically solvable Schr\"odinger potential for
the graviton fluctuation modes, we have not addressed the origin of the supergravity Lagrangian that
would result in the BPS domain wall solution whose  metric leads to the proposed Schr\"odinger
potential. Certainly, this is an important outstanding issue.

The analytic solvability of the model for the graviton  fluctuations lends itself to the  further
study of fermionic and spin-one fluctuation modes. In addition, while we chose to study the
fluctuation modes for co-dimension one (domain wall) configuration in five-dimensions, the study to
other dimensions is readily available and is relegated to future work.

\vskip0.4cm
 \emph{Acknowledgements:} We thank Valery Romanovski for useful discussions. This research
was supported by the Senior Scientist Award  (M.C.) and by a research 
grant  of the Slovenian Research Agency (M.R.) and
by the Ministry of Science and Technology of Slovenia (M.R.), and  in
part by Fay R. and Eugene L. Langberg endowed Chair  (M.C.) and Department 
of Energy Grant EY-76-02-3071 (M.C.). M.C. would like to thank The Center 
for Applied Mathematics and Theoretical Physics for hospitality. 

%\newpage

%%%%%%%%%%%%%%%%%%%%%%%%%%%%%%%%%%%%%%%%%%%%%%%%%%%%%%%%%%%%%%%%%
%%%
%%%                     BIBLIOGRAPHY
%%%
%%%%%%%%%%%%%%%%%%%%%%%%%%%%%%%%%%%%%%%%%%%%%%%%%%%%%%%%%%%%%%%%%

%\newpage
%\vskip .75 in

\end{document}